\documentclass[%
reprint,
 amsmath,amssymb,
 aps,
 pre,
]{revtex4-2}

\usepackage{graphicx}
\usepackage{bm}
\usepackage{xcolor}
\usepackage{tikz}
\usetikzlibrary{arrows.meta}

\newcommand{\Rey}{\mathrm{Re}}
\newcommand{\We}{\mathrm{We}}
\newcommand{\Oh}{\mathrm{Oh}}
\newcommand{\Ca}{\mathrm{Ca}}

\begin{document}

\preprint{APS/123-QED}

\title{Unifying Viscocapillary and Inertial Regimes in Selective Withdrawal\\} 

\author{Sabbir Hassan}
  \email{sabbir.hassan@ttu.edu}
\author{Sukalyan Bhattacharya}%
\author{Arsalan Abutalebi}%
\author{Gordon F. Christopher}%
\affiliation{%
 Department of Mechanical Engineering\\
 Texas Tech University 
}%


\author{Darryl James}
\affiliation{
 Department of Mechanical Engineering\\
 University of South Alabama
}%


\date{\today}

\begin{abstract}
Selective withdrawal extracts a single fluid phase from a multilayer liquid stratified under gravity. During this process, entrainment occurs when more than one fluid is withdrawn, often inadvertently. Existing studies identify conditions for selective withdrawal and provide scaling laws within distinct limiting regimes, including viscocapillary-dominated entrainment at low Reynolds number and inertia-dominated entrainment at high Reynolds number. However, a single unifying representation across these limits remains to be established, particularly under transitional conditions and when the lower layer exhibits non-Newtonian rheology. Selective-withdrawal experiments spanning these conditions were conducted using a Newtonian upper layer of PDMS or soybean oil and a lower layer consisting of either Newtonian RO water or a shear-thinning xanthan-gum solution. The present measurements are combined with representative literature data to develop a unified framework connecting these previously separated regimes through a ``Moody-diagram''-type representation for selective withdrawal. The normalized critical submergence height is organized using a Reynolds-like control parameter, while surface-tension effects enter through the capillary length embedded in the lateral scaling length. The resulting master curve captures the transition between dominant balances, connecting viscous- and shear-influenced entrainment to inertia-dominated entrainment. Quantitative regression confirms that the low-Reynolds-number branch closely follows the theoretically predicted $-1/3$ scaling, whereas the high-Reynolds-number branch approaches a weakly varying inertial plateau. The same asymptotic behavior is also recovered when the measured critical submergence height is examined directly using the corresponding dimensional viscous and inertial scales, providing an independent validation of the normalized collapse. The results further indicate that shear thinning primarily modifies the viscous contribution through a shear-rate-dependent effective viscosity, while the inertial baseline scale that anchors the normalization remains unchanged over the conditions examined. This regime-spanning representation avoids regime-by-regime correlation switching and provides a compact, physically interpretable framework for comparing entrainment thresholds across the Newtonian and shear-thinning systems examined here.
\end{abstract}

\maketitle

\section{Introduction}
Withdrawing fluid from a stratified system can compromise phase purity. Selective withdrawal provides a generic solution to this problem. In this configuration (Fig.~\ref{fig:setup_simple}), a withdrawal tube is positioned above an initially flat liquid--liquid interface. Increasing the withdrawal rate deforms the interface into a hump (Fig.~\ref{fig:entrainment_qualitative}).  
Lowering the tube produces the same qualitative deformation sequence. The hump increases in amplitude while its curvature localizes near the apex. At a critical condition, the apex sharpens and a lower-layer tendril forms. The lower layer is then entrained into the tube, and the system transitions from single-phase to two-phase withdrawal. This transition underpins stratified multiphase transport and separation, from emulsion management to industrial withdrawal operations.  
It also motivates Strategic Petroleum Reserve withdrawal scenarios, where phase purity is required near throughput constraints \cite{hartenberger2011, sabbir2022}.

\begin{figure}[htpb]
\centering
\begin{tikzpicture}[x=1cm,y=1cm,>=Latex,scale=0.9]
  \draw[thick] (-2.6,-1.5) rectangle (2.6,1.55);
  \draw[thick] (-2.6,0) -- (2.6,0);
  \node[anchor=west] at (-2.45,0.80) {Upper fluid};
  \node[anchor=west] at (-2.45,-0.85) {Lower fluid};
  \node[anchor=east] at (2.45,-0.18) {Interface};
  \fill[black] (-0.38,1.55) rectangle (0.38,0.88);
  \draw[thick] (-0.38,1.55) -- (-0.38,0.88) -- (0.38,0.88) -- (0.38,1.55);
  \draw[->,thick] (0,1.18) -- (0,1.85) node[above] {$Q$};
  \draw[<->,thick] (1.15,0.04) -- (1.15,0.88);
  \node[anchor=west] at (1.25,0.46) {$S$};
\end{tikzpicture}
\caption{Simplified schematic of selective withdrawal from a two-layer liquid system. A withdrawal tube is positioned a distance $S$ above the initially flat interface and removes fluid from the upper layer at volumetric flow rate $Q$. At the critical condition, $S=S_c$, further deformation leads to entrainment of the lower layer.}
\label{fig:setup_simple}
\end{figure}

A persistent challenge is that entrainment onset lacks a universal scaling formulation. Instead, different force balances dominate across parameter space.  
In the low-$\Rey$ viscocapillary limit, viscous stresses and capillarity govern entrainment. Classical studies show that capillary-number scalings capture thresholds and morphology evolution \cite{cohen2002, cohen2004scaling, blanchette2009}.  
In high-$\Rey$ regimes, inertia competes with gravity and surface tension.  
Recent work therefore emphasizes regime maps and correlations using $\We$ and $\Oh$ \cite{sabbir2022, lubin}. Each framework is compelling within its stated validity range. However, a smooth and interpretable connection between these limits remains missing. The gap becomes more apparent when the lower layer is non-Newtonian.  
A unified representation that spans regimes is therefore still needed.

The contribution sought here is not a new limiting exponent in either asymptotic regime. Those limits are already supported by the viscocapillary and inertial literature. The new element is to test whether the same normalized response can connect those limits on common axes, quantify the crossover without switching correlations, and retain the same primary organization when the lower layer is shear thinning. The present experiments are therefore analyzed together with literature data that extend the accessible viscosity, density, flow-rate, and tube-size ranges.

This gap is not merely aesthetic; it is also practical for modeling. Regime-specific laws naturally encourage piecewise reasoning in applications. One typically selects a regime and then applies its correlation. Real experiments often occupy transitional conditions between classical limits. Operating systems similarly encounter states where neither limit is sufficient. Shear-thinning fluids add ambiguity to viscous stress estimates. The relevant viscosity depends on the local interfacial shear rate. Consequently, Newtonian scalings may appear to fail for shear-thinning cases. Such deviations can arise from inconsistent shear-rate definitions. They can also reflect inconsistent coupling to kinematics and local geometry.

The goal of this paper is a regime-spanning organization of entrainment onset. It is intended to unify viscocapillary and inertial limits on common axes while retaining explicit roles for inertia, viscosity, gravity, and capillarity.  
A master curve analogous to the Moody diagram in pipe flow is introduced.  
Rather than friction factor versus Reynolds number, normalized critical submergence is plotted against a Reynolds-like control parameter.  
Surface-tension effects enter through the capillary length.  
In this representation, viscous and inertial regimes appear as limiting behaviors of a single curve rather than as separate scaling laws.

\subsection{Experimental setup and material characterization}

A withdrawal tube was positioned above a two-layer liquid interface in a rectangular acrylic tank \cite{Hassan2024Dissertation}. The tank measured $30.48~\mathrm{cm}$ in depth, $60.96~\mathrm{cm}$ in length, and $30.48~\mathrm{cm}$ in width, and the withdrawal tube had an inner diameter of $d=0.53~\mathrm{cm}$. The tube was aligned normal to the initially flat interface and positioned near the center of the withdrawal section, more than $15~\mathrm{cm}$ from the surrounding tank walls, to minimize
confinement and wall effects \cite{cohen2004scaling,Hassan2024Dissertation}.

Six fluid combinations were examined. The withdrawn upper layer consisted of either nominal 64 cSt soybean oil or nominal 5 cSt polydimethylsiloxane (PDMS) silicone oil. The lower layer consisted of reverse-osmosis (RO) water, used as the Newtonian reference fluid, or an aqueous xanthan-gum solution at a concentration of either 300 or
600 mg/L. Thus, each Newtonian upper fluid was paired with one Newtonian lower fluid and two shear-thinning lower fluids. The densities, upper-fluid viscosities, lower-layer rheological parameters, and interfacial tensions used in the analysis are summarized in Table~\ref{tab:props}.

\begin{table*}[!t]
\caption{Fluid combinations and representative material properties used in
the selective-withdrawal experiments \cite{Hassan2024Dissertation}. PDMS denotes polydimethylsiloxane, Xn--H$_2$O denotes an aqueous xanthan-gum solution, and RO water denotes reverse-osmosis water. The designations 64 cSt and 5 cSt identify the nominal kinematic-viscosity grades of the upper fluids, whereas $\mu_U$ denotes the measured dynamic viscosity used in the analysis. Here, $\rho_U$ and $\rho_L$ are the upper- and lower-fluid densities, respectively, and $\gamma$ is the interfacial tension. For the
shear-thinning xanthan-gum solutions, the measured viscosity is represented by $\eta_L=m\dot{\gamma}^{\,n-1}$, where $m$ is the consistency index, $n$ is the dimensionless flow-behavior index, and $\eta_{\mathrm{pl}}$ is the low-shear viscosity plateau. The xanthan-gum parameters correspond to representative measurements at $21^{\circ}\mathrm{C}$; dashes indicate quantities not applicable to Newtonian RO water.}
\label{tab:props}
\centering
\scriptsize
\setlength{\tabcolsep}{2.6pt}
\renewcommand{\arraystretch}{1.18}
\begin{ruledtabular}
\begin{tabular}{llccclcccc}
Upper fluid &
Lower fluid &
$\rho_U$ &
$\mu_U$ &
$\rho_L$ &
\multicolumn{4}{c}{Lower-layer rheology} &
$\gamma$ \\
&
&
(kg/m$^3$) &
(cP) &
(kg/m$^3$) &
Model &
$m$ (mPa\,s$^n$) &
$n$ &
$\eta_{\mathrm{pl}}$ (cP) &
(N/m) \\
\hline
Soybean oil (64 cSt) & RO water                & 863 & 55   & 990    & Newtonian & --    & --   & --    & 0.028 \\
Soybean oil (64 cSt) & 300 mg/L Xn--H$_2$O    & 863 & 55   & 990.25 & Power law & 27.11 & 0.62 & 19.0  & 0.028 \\
Soybean oil (64 cSt) & 600 mg/L Xn--H$_2$O    & 863 & 55   & 990.5  & Power law & 33.12 & 0.64 & 22.34 & 0.028 \\
PDMS oil (5 cSt)     & RO water                & 913 & 4.35 & 990    & Newtonian & --    & --   & --    & 0.038 \\
PDMS oil (5 cSt)     & 300 mg/L Xn--H$_2$O    & 913 & 4.35 & 990.25 & Power law & 27.11 & 0.62 & 19.0  & 0.038 \\
PDMS oil (5 cSt)     & 600 mg/L Xn--H$_2$O    & 913 & 4.35 & 990.5  & Power law & 33.12 & 0.64 & 22.34 & 0.038 \\
\end{tabular}
\end{ruledtabular}
\normalsize
\end{table*}

The tank was divided into withdrawal and return-flow sections by an acrylic
partition so that fluid discharged from the circulation loop did not
directly disturb the interface beneath the withdrawal tube. The required
return-flow conditioning depended on the upper fluid because the two
withdrawn liquids covered substantially different viscosity and flow-rate
ranges. For the higher-viscosity soybean-oil systems, the relatively low
withdrawal rates produced a low-momentum return stream that did not
noticeably perturb the interface. In contrast, the lower-viscosity PDMS
systems permitted substantially higher withdrawal rates and consequently
produced a higher-momentum return jet. A 3D-printed PLA baffle was therefore
installed in the return-flow region during the PDMS experiments to dissipate
the jet momentum before the recirculated liquid could influence the
measurement region.

Several checks were performed to assess sensitivity of the measured entrainment threshold to the finite dimensions of the apparatus. The withdrawal tube was displaced laterally by approximately two tube diameters
from its nominal centered position, with no measurable change in the critical submergence height. The thickness of the lower aqueous layer was
also varied between approximately $2.5$ and $7.6~\mathrm{cm}$ ($1$--$3~\mathrm{in.}$), again without a measurable change in the critical submergence height. These checks indicate that, over the conditions examined, the entrainment measurements were not appreciably affected by
tank-wall proximity or lower-layer depth.

\begin{figure*}[htpb]
\centering
\includegraphics[width=0.78\textwidth]{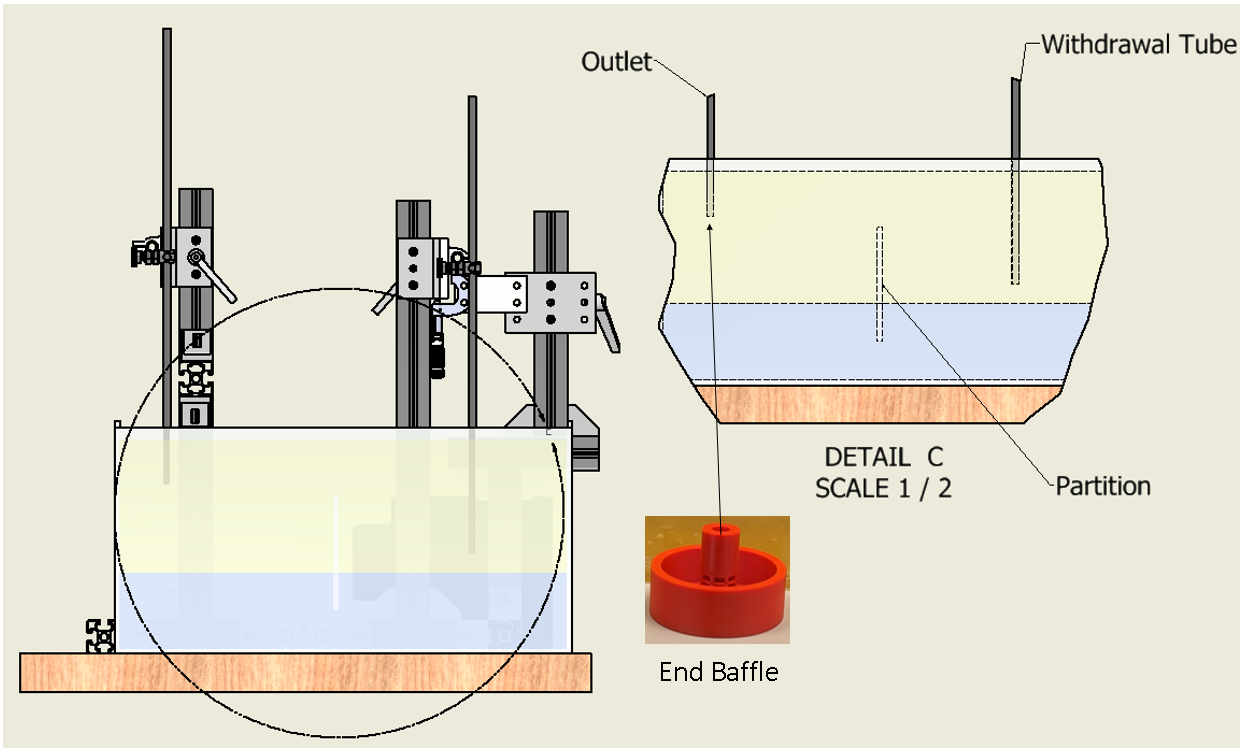}
\caption{Side view of the modeled selective-withdrawal apparatus. The withdrawal
tube is positioned normal to the initially flat liquid--liquid interface,
while the central partition separates the withdrawal region from the
return-flow region. Cameras and flexible tubing are omitted for clarity.
The experimental configuration is documented in
Ref.~\cite{Hassan2024Dissertation}.}
\label{fig:experimental_setup}
\end{figure*}

For each imposed flow rate $Q$, the initially flat interface was used to
establish the reference position of the withdrawal tube. After the flow
rate had stabilized, the tube was moved quasi-statically toward the interface. The tube was initially lowered in increments of $0.1~\mathrm{in.}$ ($2.54~\mathrm{mm}$); near the transition to entrainment, the increment was reduced to $0.05~\mathrm{in.}$ ($1.27~\mathrm{mm}$) \cite{Hassan2024Dissertation}. At the first observable onset of lower-layer entrainment, the pump was stopped and the critical submergence height $S_c$ and corresponding flow rate were recorded. Repeated measurements were obtained at identical flow conditions to assess experimental repeatability. Temperature was monitored throughout testing, and measurements were suspended when temperature changes were sufficient to affect the controlled fluid properties \cite{Hassan2024Dissertation}. The resulting protocol provided a quasi-static measure of the entrainment threshold while reducing  
to transient overshoot and approach history.

For each fluid-pair/flow-rate condition, repeated measurements of \(S_c\) were summarized by the sample mean and a 95\% Student-\(t\) confidence interval,
\begin{equation}
\bar{S}_c \pm t_{0.975,n-1}\frac{s}{\sqrt{n}},
\label{eq:ScCI}
\end{equation}
where \(s\) is the sample standard deviation and \(n\) is the number of repeats. Across the 24 present-study conditions, the relative half-width of this interval ranged from approximately 0.2\% to 3.6\% \cite{Hassan2024Dissertation}. These uncertainties are small compared with the regime-scale variation in the collapsed response and are retained as experimental-repeatability estimates rather than absorbed into the fitted scaling exponents.

The rheological behavior of the fluids listed in Table~\ref{tab:props} was characterized independently before the selective-withdrawal experiments. Viscosity measurements were performed using a Discovery HR20 rheometer (TA Instruments) equipped with a stainless-steel double-wall concentric-cylinder geometry. Flow sweeps were conducted over shear rates from $0.01$ to $300~\mathrm{s^{-1}}$ \cite{Hassan2024Dissertation}. The viscosities of soybean oil and PDMS remained approximately independent of shear rate at fixed temperature, supporting their treatment as Newtonian upper-layer fluids. At approximately $21^{\circ}\mathrm{C}$, representative dynamic viscosities were approximately $55~\mathrm{cP}$ for soybean oil and
$4.8~\mathrm{cP}$ for PDMS; the values used in the scaling analysis are
reported in Table~\ref{tab:props}.

In contrast, the 300 and 600 mg/L xanthan-gum solutions exhibited
shear-thinning behavior over the measured shear-rate range together with a
low-shear viscosity plateau. Over the shear-thinning region, the lower-layer viscosity was represented by
\begin{equation}
\eta_L(\dot{\gamma})=m\dot{\gamma}^{\,n-1},
\label{eq:powerlaw}
\end{equation}
where $m$ is the consistency index and $n$ is the flow-behavior index. At $21^{\circ}\mathrm{C}$, the 300 mg/L solution was described by $m=27.11~\mathrm{mPa\,s^n}$ and $n=0.62$ over $2.55<\dot{\gamma}<300~\mathrm{s^{-1}}$, with a low-shear plateau of approximately $19.0~\mathrm{mPa\,s}$. The 600 mg/L solution was described by $m=33.12~\mathrm{mPa\,s^n}$ and $n=0.64$ over $2.98<\dot{\gamma}<300~\mathrm{s^{-1}}$, with a low-shear plateau of approximately $22.34~\mathrm{mPa\,s}$ \cite{Hassan2024Dissertation}. The measured rheological trends and representative constitutive parameters are summarized in Table~\ref{tab:props}; the complete viscosity profiles are reported in
Ref.~\cite{Hassan2024Dissertation}.

Interfacial tension was measured independently using pendant-drop
tensiometry. Three measurements were obtained for each fluid combination,
and the recorded drop profiles were analyzed using OpenDrop
\cite{Hassan2024Dissertation}. Representative interfacial tensions were
approximately $0.028~\mathrm{N\,m^{-1}}$ for the soybean-oil systems and
$0.038~\mathrm{N\,m^{-1}}$ for the PDMS systems.

\subsection{Regime map}
\begin{figure*}[!t]
\centering
\includegraphics[width=160mm]{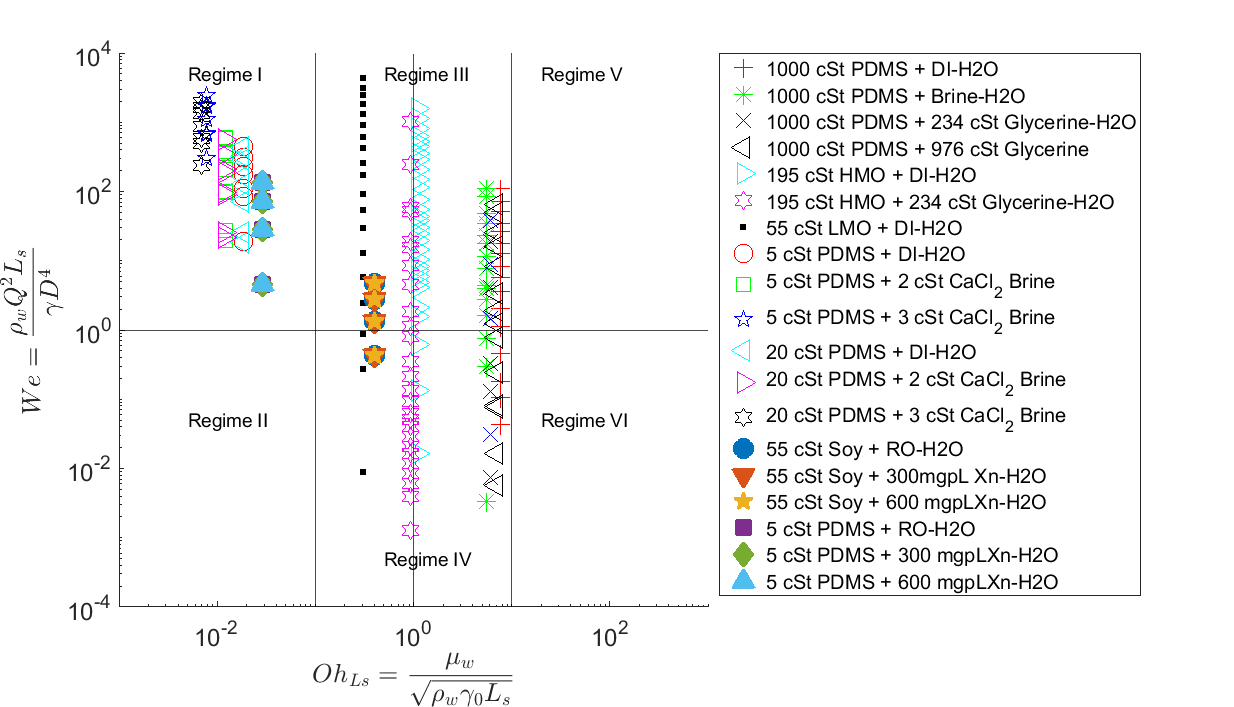}
\caption{Regime map (Weber--Ohnesorge) contextualizing the present experiments relative to viscocapillary (Cohen-type) conditions and inertial (Hassan-type) conditions. The distribution of cases motivates the need for a unified, regime-spanning representation of entrainment onset.}
\label{fig:regimemap}
\end{figure*}

Operating conditions are first visualized using the Weber--Ohnesorge plane in Fig.~\ref{fig:regimemap}. The inertial-mapping framework introduced previously is used for this purpose \cite{sabbir2022}. The aim is not a universal collapse of the entrainment threshold. Instead, the plot serves as a diagnostic for regime proximity. It shows overlap with Cohen-type viscocapillary environments and inertial environments. This overlap motivates a regime-spanning representation for entrainment onset. Such a representation should not presuppose a single dominant balance.

\begin{table*}[htpb]
\caption{Approximate nondimensional-parameter ranges used to contextualize the present dataset relative to representative viscocapillary (Cohen) and inertial (Sabbir et al.) literature. The tabulated tube-\(\Rey\), \(\We\), and \(\Oh\) ranges follow the definitions used in the cited source studies and are used only for regime context; the master-curve coordinate \(\Rey_{L_s}\) is recomputed consistently from Eq.~(\ref{eq:ReLs}). ``Tube insertion'' indicates whether the withdrawal tube penetrates the interface; all datasets shown here correspond to no insertion.}
\label{tab:regimeranges}
\centering
\setlength{\tabcolsep}{8pt}
\renewcommand{\arraystretch}{1.15}
\begin{tabular}{l c c c c}
\hline
Literature & Tube $\Rey$ range & $\We$ range & $\Oh$ range & Tube insertion \\
\hline
Cohen (2004) & 0.01--200 & 0.001--4300 & 0.3--7 & No \\
Sabbir et al.\ (2022) & 1400--34000 & 18--2500 & 0.005--0.02 & No \\
Current study & 0.9--108 & 0.4--134 & 0.03--0.4 & No \\
\hline
\end{tabular}
\end{table*}

Table~\ref{tab:regimeranges} makes the modeling requirement explicit.  
Cohen-type data span very low-to-moderate Reynolds numbers and larger $\Oh$.
These conditions emphasize viscocapillary control of entrainment and morphology. In contrast, Sabbir et al.\ data occupy very high Reynolds numbers and small $\Oh$. Those conditions emphasize inertia competing with gravity and surface tension. The present experiments deliberately sit between these extremes. The Reynolds numbers of the present experiments are low-to-moderate, while $\Oh$ is smaller than the range reported by Cohen.  
The corresponding Weber numbers also enter ranges where inertia is not negligible. Thus, the dataset samples the ``connecting tissue'' between classical limits. It therefore provides an ideal testbed for a unified representation.

Figure~\ref{fig:regimemap} reinforces the same interpretation. A purely $\Ca$-based viscocapillary collapse is not expected to be robust here. Such a collapse could suffice at uniformly low $\Rey$ and large $\Oh$. In the present data, inertia influences the threshold at higher flow rates. The effect is strongest for lower-viscosity upper layers. A unifying description must therefore accommodate both limits. It must also remain interpretable across transitional conditions. This is the purpose of the master curve in Eq.~(\ref{eq:master}). Inertia supplies a baseline scale through $S_I$. Viscous and shear physics then contribute a predictable correction. That correction is quantified by $\Rey_{L_s}^{-1/3}$.

\subsection{Roadmap of the paper}
The paper is organized around four major sections. Section~II develops the scaling framework, the two asymptotic submergence scales, and the composite nondimensional representation. Section~III presents the collapse, quantifies both asymptotic branches, and tests the same scalings in dimensional variables. The roles of the lateral scale, surface tension, and shear thinning are then discussed together with the limits of the available data.  
Section~IV summarizes the regime-spanning picture and its implications.

\section{Scaling framework and composite model}

A consistent starting point for a regime-spanning organization is the interfacial normal-stress balance.  
At the interface, the jump in normal stress must be supported by capillary pressure associated with curvature.  
In compact form,
\begin{eqnarray}
\hat{\bm{n}}\cdot(\bm{T}_U-\bm{T}_L)\cdot\hat{\bm{n}}
=
\gamma\,(\nabla\cdot\hat{\bm{n}}),
\label{eq:normalstress}
\end{eqnarray}

\noindent where $\hat{\bm{n}}$ is the unit normal to the interface and $\bm{T}_{U,L}$ are the stress tensors in the upper and lower fluids.  
Equation~(\ref{eq:normalstress}) is general and does not presuppose a particular regime.  
Regime dependence enters only through how the stress contributions and curvature are estimated.  
In practice, these magnitudes are expressed in terms of $Q$, $S_c$, and a horizontal scaling length.  
That horizontal length represents the lateral region over which suction and curvature are sampled by the interface.

In the present experiments, the upper layer is Newtonian.  
The lower layer is either Newtonian water or a shear-thinning xanthan-gum solution.  
For shear-thinning cases, a generalized-Newtonian description captures the leading effect on entrainment.  
The dominant consequence is a viscosity that varies with the local shear rate.  
The lower-layer viscosity is represented by the power-law form
\begin{eqnarray} \nonumber
\eta_L(\dot{\gamma}) = m\,\dot{\gamma}^{\,n-1},
\qquad (n<1\ \text{for shear thinning}),
\label{eq:powerlaw2}
\end{eqnarray}

\noindent where $m$ and $n$ are obtained from rheometry over the shear-rate window relevant near the interface.  
Importantly, the structure of the stress balance is unchanged across fluid pairs.  
What changes is the mapping from flow and geometry to the viscosity scale entering viscous stresses.  
That mapping determines which effective viscosity is appropriate for a given operating condition.

Gravity and surface tension jointly influence interfacial curvature.  
A central geometric scale is therefore the capillary length,
\begin{eqnarray}
l_c=\sqrt{\frac{\gamma}{\Delta\rho g}},
\qquad
\Delta\rho=\rho_L-\rho_U.
\label{eq:lc}
\end{eqnarray}
The capillary length links near-interface curvature to far-field hydrostatics.  
To position the present experiments relative to the broader literature, $\We$ and $\Oh$ are used \cite{sabbir2022}:
\begin{eqnarray}
\We &=& \frac{\rho_w U_0^2 l_c}{\gamma},
\label{eq:we}
\\
\Oh &=& \frac{\mu_w}{(\rho_w \gamma l_c)^{1/2}}.
\label{eq:oh}
\end{eqnarray}
Here $U_0$ is a characteristic suction-driven velocity scale; in the scaling developed below it is estimated as \(U_0\sim Q/S_c^2\). The quantities \((\rho_w,\mu_w)\) denote the density and viscosity scale of the withdrawn (upper) fluid \cite{Hassan2024Dissertation}.  
In the classical viscocapillary limit, entrainment is commonly organized by a capillary number,
\begin{eqnarray}
\Ca \sim \frac{\mu_w U}{\gamma}.
\label{eq:ca}
\end{eqnarray}
A purely $\Ca$-based framework implicitly neglects inertia in the stress balance.  
This assumption is intentionally violated for portions of the present dataset.  
This motivates introducing a Reynolds-like control parameter for regime-spanning unification.

\subsection{Two asymptotic geometries}
\begin{figure*}[!t]
\centering
\includegraphics[width=160mm]{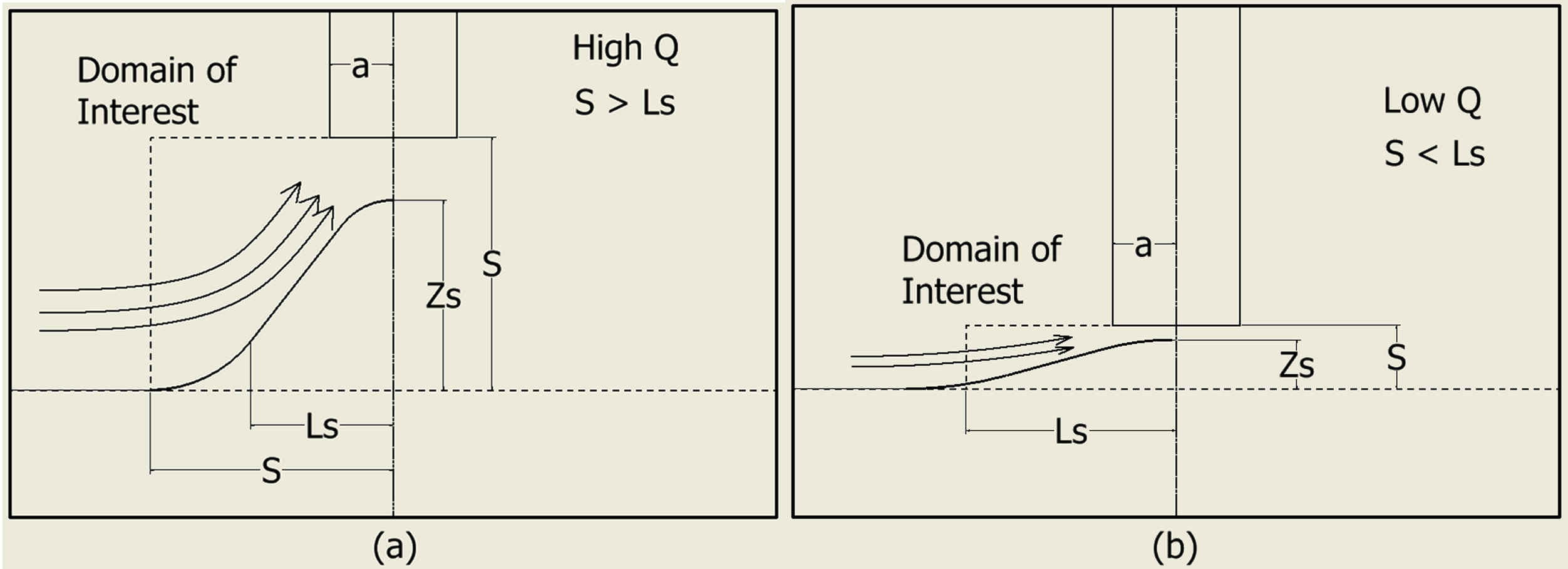}
\caption{Schematic scaling pictures motivating two asymptotic balances: (a) inertia-dominated entrainment where the vertical scale is set primarily by inertia and gravity ($S_c \gtrsim L_s$); (b) viscous/shear-dominated entrainment where shear stresses provide a leading correction over a horizontal scaling length ($S_c \lesssim L_s$).}
\label{fig:sketch}
\end{figure*}

The scaling analysis highlights two asymptotic geometric pictures \cite{Hassan2024Dissertation}.  
The relevant geometric picture depends on withdrawal strength and tube proximity.  
These limits are summarized by the sketch in Fig.~\ref{fig:sketch}.  
In both cases, entrainment is set by how suction-driven flow couples to the interface.  
That coupling involves a vertical deformation scale, $S_c$.  
It also involves a horizontal scaling length, $L_s$.  
In one limit, inertia drives deformation against gravity, yielding an inertial scale for $S_c$.  
In the other limit, viscous and shear stresses dominate or contribute a strong correction.  
This produces a different scaling exponent for the entrainment threshold.  
A unified representation must recover both limiting behaviors.  
It must also capture the continuous transition between them.

The analysis yields two leading-order submergence scales \cite{Hassan2024Dissertation}.  
First, an inertial--gravitational balance gives
\begin{eqnarray}
S_I \sim \left(\frac{Q^2}{g}\right)^{1/5}.
\label{eq:SI}
\end{eqnarray}
This balance compares suction-driven inertial stresses, which grow with $Q$, to hydrostatic restoration set by $g$.  
Equation~(\ref{eq:SI}) therefore provides a baseline vertical length scale.  
It remains relevant even when viscous and capillary effects contribute appreciably.  
Inertia and gravity are present throughout the parameter space considered here.

Second, viscous and shear stresses can provide a significant correction \cite{Hassan2024Dissertation}.  
In that case, the analysis yields the viscous correction scale
\begin{eqnarray}
S_V \sim \left(\frac{\mu_w Q}{\Delta\rho\, g\, L_s}\right)^{1/3},
\label{eq:SV}
\end{eqnarray}
where $L_s$ is a horizontal scaling length.  
This length reflects the lateral region over which curvature and shear are imposed by the withdrawal flow.  
It depends on both the near-field suction geometry and the gravity--capillary curvature scale.  
A practical crossover choice is \cite{Hassan2024Dissertation}:
\begin{eqnarray}
L_s \approx \max\!\left(l_c,\; a\right),
\qquad
a=\frac{d}{2}.
\label{eq:LsChoice}
\end{eqnarray}
This definition is a scaling prescription rather than a constitutive law. It selects the larger of the gravity--capillary curvature scale and the nozzle radius, thereby preventing the sampled lateral scale from becoming smaller than either physically imposed lower bound. When \(l_c>a\), gravity-capillary curvature sets the scale and \(L_s\rightarrow l_c\); when \(a>l_c\), near-nozzle geometry sets the scale and \(L_s\rightarrow a\). The \(\max\) function is continuous, although not differentiable at \(a=l_c\); this corner does not enter the dynamics because \(L_s\) is used only as a characteristic scale and is never differentiated. Both orderings occur in the data represented by the master plot. The composite definition is therefore exercised by the dataset rather than introduced only as a formal limiting argument.

For regime-spanning comparison, $S_c$ is normalized by the inertial scale $S_I$ and interpreted as a response variable analogous to a friction factor.
The normalized response is defined as
\begin{eqnarray}
Y \;\equiv\; \frac{S_c}{S_I}
\;=\;
\frac{S_c\,g^{1/5}}{Q^{2/5}}.
\label{eq:Ydef}
\end{eqnarray}
If inertia and gravity alone set the entrainment threshold, $Y$ approaches a constant. Departures from that constant quantify viscous, shear, and capillary corrections.

A Reynolds-like control parameter based on $L_s$ is defined as
\begin{eqnarray}
\Rey_{L_s}
\;\equiv\;
\frac{\Delta\rho\, g^{2/5}\, Q^{1/5}\, L_s}{\mu_w}.
\label{eq:ReLs}
\end{eqnarray}
This group increases with flow rate and decreases with viscosity. The factor $Q^{1/5}$ follows directly from the inertial scaling. It sets the natural velocity and length scales used in the unification. The regime-spanning hypothesis is expressed as the lowest-order composite interpolation
\begin{eqnarray}
Y \;=\; A \;+\; B\,\Rey_{L_s}^{-1/3}.
\label{eq:master}
\end{eqnarray}
Equation~(\ref{eq:master}) is not asserted to be an exact closed-form solution of the full free-boundary problem. Rather, it is a compact composite closure constrained to recover the two asymptotic balances: an inertial plateau at large \(\Rey_{L_s}\), where \(Y\rightarrow A\), and a viscous correction proportional to \(\Rey_{L_s}^{-1/3}\) as \(\Rey_{L_s}\) decreases.

Substituting Eqs.~(\ref{eq:Ydef}) and (\ref{eq:ReLs}) into Eq.~(\ref{eq:master}) gives the equivalent dimensional representation
\begin{eqnarray}
S_c
&=&
A\left(\frac{Q^2}{g}\right)^{1/5}
+
B\left(\frac{\mu_w Q}{\Delta\rho\,g\,L_s}\right)^{1/3}
\nonumber\\
&=& A\,S_I+B\,S_V.
\label{eq:dimensional_master}
\end{eqnarray}
This form makes the physical content of the interpolation explicit: the first term is the inertial--gravitational scale and the second is the viscous correction scale.

Because \(Q\) appears in both \(Y\) and \(\Rey_{L_s}\), the normalized master plot should be interpreted as a similarity representation rather than as an ordinary regression between statistically independent predictor and response variables. The shared occurrence of \(Q\) is a consequence of the nondimensionalization and can strengthen apparent graphical correlation if considered by itself. For this reason, the quantitative assessment below uses the normalized regressions to test the predicted asymptotic exponents and, separately, tests \(S_c\propto S_V\) and \(S_c\propto S_I\) directly in dimensional variables. Agreement in dimensional space provides a check that the observed regime structure is not solely an artifact of shared-variable normalization.

For shear-thinning lower layers, viscosity must be evaluated at an appropriate shear-rate scale. A physically motivated estimate uses a characteristic velocity divided by a characteristic length.  
Using the suction-driven velocity scale $U\sim Q/S_c^2$, the effective shear-rate scale is estimated as
\begin{eqnarray}
\dot{\gamma}_{\text{eff}}
\sim
\frac{U}{L_s}
\sim
\frac{Q/S_c^2}{L_s},
\label{eq:gammaeff}
\end{eqnarray}
The corresponding effective viscosity is defined as
\begin{eqnarray}
\eta_{L,\text{eff}} = m\,\dot{\gamma}_{\text{eff}}^{\,n-1}.
\label{eq:etaeff}
\end{eqnarray}
A distinction is required between the viscosity used in the primary Reynolds-like coordinate and the rheological correction associated with the lower layer.  
In $\Rey_{L_s}$, $\mu_w$ represents the viscosity scale of the withdrawn Newtonian upper fluid.  
The generalized-Newtonian contribution of the lower layer enters the nondimensional viscous stress through the factor
\begin{eqnarray}
\Lambda_{\mathrm{NN}}
=
\frac{m_L}{\mu_w}
\left(
\frac{Q^{1/5}g^{2/5}}{L_s}
\right)^{n_L-1}.
\label{eq:LambdaNN}
\end{eqnarray}
For $n_L=1$, this factor reduces to the corresponding Newtonian viscosity-ratio contribution.  
For $n_L<1$, the contribution decreases with increasing characteristic shear rate, reflecting the reduction in apparent viscosity associated with shear thinning.  
The generalized-Newtonian lower layer therefore modifies the viscous stress correction without requiring a different inertial normalization.  
The effective viscosity in Eq.~(\ref{eq:etaeff}) provides an equivalent local interpretation of this shear-rate-dependent correction.

\section{Results and quantitative validation}

\subsection{Qualitative interface evolution}
\begin{figure*}[!t]
\centering
\includegraphics[width=160mm]{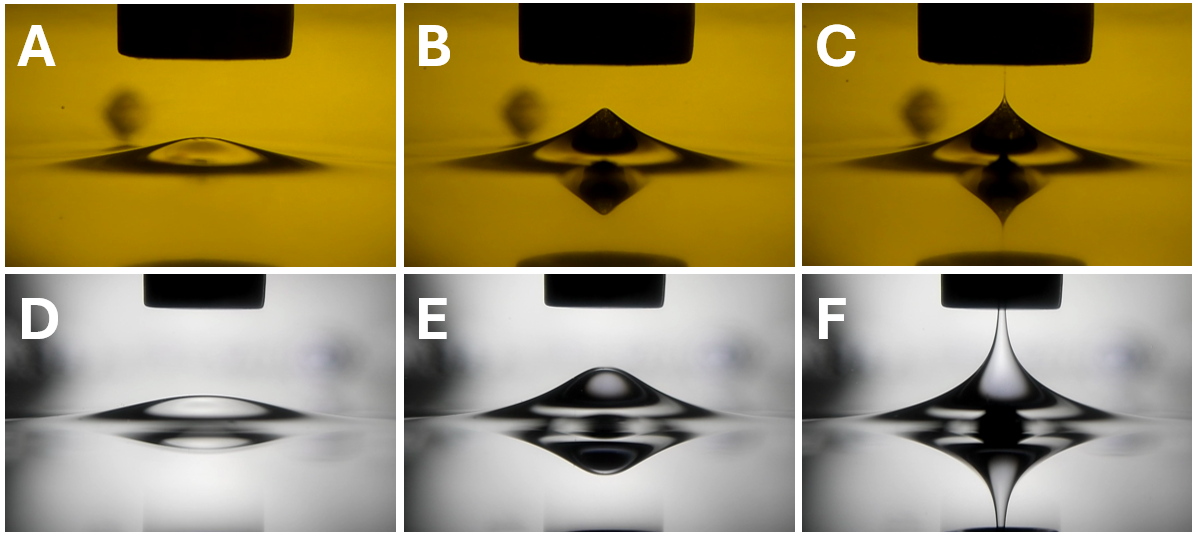}
\caption{Progressive deformation of the liquid--liquid interface during selective withdrawal for representative Newtonian lower-layer cases. The top row (A--C) corresponds to soybean oil over RO water and the bottom row (D--F) to PDMS oil over RO water. The images show initial deformation, intermediate hump growth, and critical/near-critical deformation immediately preceding entrainment. The soybean-oil system develops a sharper and more localized peak than the PDMS system at comparable stages. Because the interface morphology depends on several fluid-property ratios and operating conditions, these images are used as a qualitative illustration of the approach to entrainment rather than as a quantitative classifier of flow regime.}
\label{fig:entrainment_qualitative}
\end{figure*}

Across all cases, the interface exhibits a reproducible sequence of deformations as the forcing is increased.  
Representative stages are shown in Fig.~\ref{fig:entrainment_qualitative}.  
At low $Q$ and larger tube heights, the response is weak and broadly distributed: a gentle, low-amplitude hump forms beneath the withdrawal tube.  
As $Q$ increases and/or the submergence decreases, the hump grows in amplitude and becomes progressively more localized, with curvature concentrating near the apex.  
Near the critical condition, the interface typically undergoes rapid apex sharpening, and a thin tendril or spout forms that connects into the withdrawal tube.  
This spout formation signals imminent entrainment of the lower layer.  
Differences in peak sharpness and localization are observed across fluid combinations and reflect the coupled influence of viscosity, interfacial tension, density contrast, and operating condition.  
The image sequence is therefore used as a qualitative illustration of the approach to entrainment rather than as a quantitative classifier of flow regime.

To construct a dataset appropriate for collapse, exclusion criteria were applied to isolate a reproducible quasi-static entrainment threshold.  
Cases affected by intermittent rag-layer interference, persistent emulsification, or ambiguous onset triggered by transients and/or recirculation were excluded \cite{Hassan2024Dissertation}.  
This filtering is motivated both scientifically and practically.  
The master curve is intended to encode deterministic stress-balance physics, and the exclusions reduce contamination by stochastic or artifact-driven variability.  
With this filtering in place, residual deviations from the master trend are more plausibly attributed to secondary physical effects not captured by the minimal model assumptions.  
Examples include subdominant capillary effects, sensitivity to the local shear-rate scale, or weak dependence on interfacial history.

\subsection{Unified collapse: the `Moody-like diagram'}
A natural baseline approach is to organize the entrainment threshold using viscocapillary parameters, most notably the capillary number $\Ca$, consistent with Cohen-type scalings \cite{cohen2004scaling, blanchette2009}.  
Within the present dataset, the soybean-oil cases overlap more strongly with conditions under which viscous stresses are prominent.  
For those cases, $\Ca$-type organization can yield a reasonable collapse and reproduces the expected qualitative trends of $S_c$ with $Q$.

However, the same viscocapillary organization is not robust across the full dataset.  
The PDMS cases extend into more transitional and inertia-influenced conditions, for which the assumption of viscous dominance becomes questionable.  
In this regime, a $\Ca$-based scaling deviates systematically, not because capillarity vanishes, but because inertia introduces an additional stress scale that $\Ca$ does not represent.  
Consequently, cases with substantial inertial contribution cannot be expected to collapse on a purely viscocapillary axis.  
This motivates a Reynolds-like organizing parameter that explicitly compares inertia and viscosity over an appropriate scaling length.

When the measured critical submergence height \(S_c\) is normalized by the inertial scale \(S_I\) and plotted against \(\Rey_{L_s}\), the compiled datasets organize into a common two-branch trend consistent with Eq.~(\ref{eq:master}). The plot is directly analogous to a Moody diagram: the vertical axis is a normalized response, \(Y\), and the horizontal axis is a Reynolds-like control parameter that organizes the transition between limiting behaviors. The graphical collapse is assessed quantitatively below rather than inferred from visual overlap alone.

\begin{figure*}[!t]
\centering
\includegraphics[width=160mm]{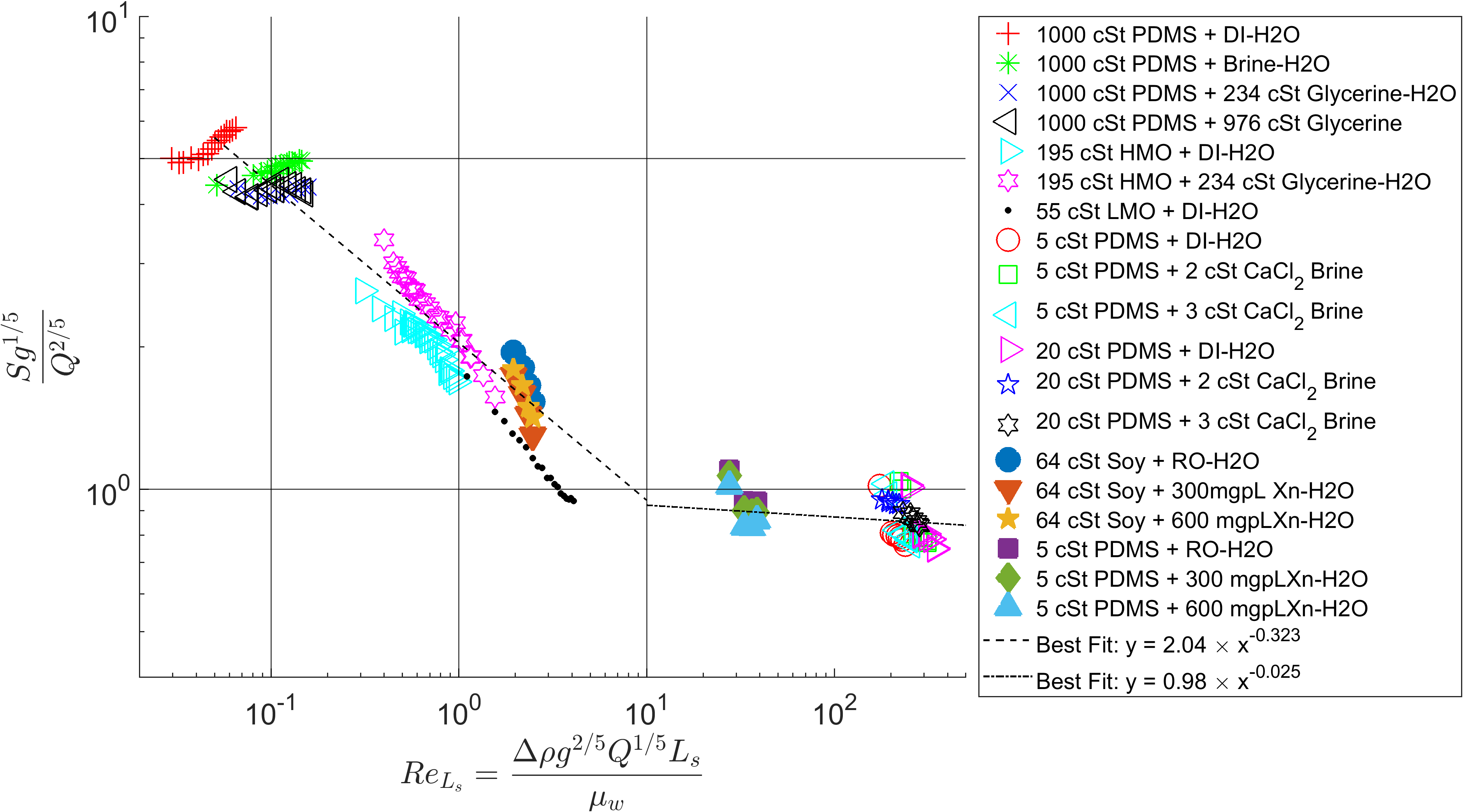}
\caption{Moody-diagram-type collapse: normalized critical submergence height \(Y=S_c g^{1/5}/Q^{2/5}\) plotted versus the Reynolds-like control parameter \(\Rey_{L_s}\). The dashed and dash-dotted guide fits retained in the figure are \(Y=2.04\,\Rey_{L_s}^{-0.323}\) and \(Y=0.98\,\Rey_{L_s}^{-0.025}\), respectively. An independent branch-wise regression of all 202 plotted points, using \(\Rey_{L_s}=10\) as a fixed crossover, is reported in Table~\ref{tab:branchfits}. The low-\(\Rey_{L_s}\) exponent is statistically consistent with \(-1/3\), whereas the high-\(\Rey_{L_s}\) dependence is much weaker and approaches an inertial plateau.}
\label{fig:moodycollapse}
\end{figure*}

\subsection{Quantitative validation of the asymptotes}

To quantify the same structure without relying on the visually selected guide lines in Fig.~\ref{fig:moodycollapse}, all 202 plotted points were reanalyzed using a fixed crossover at \(\Rey_{L_s}=10\). The compiled set contains 129 Cohen points, 49 inertial literature points, and 24 present-study condition means \cite{cohen2004scaling,sabbir2022,Hassan2024Dissertation}. The present experiments enter the regression as condition means rather than individual repeats, so conditions with more replicate measurements do not receive disproportionate statistical weight. In each region, a free power law,
\begin{equation}
Y=C\,\Rey_{L_s}^{\,p},
\label{eq:branchfit}
\end{equation}
was fit by nonlinear least squares. The cutoff is used only to quantify the two visually distinct asymptotes; it is not introduced into Eq.~(\ref{eq:master}) as a discontinuous regime switch.

\begin{table*}[!t]
\caption{Branch-wise regression of the data shown in Fig.~\ref{fig:moodycollapse}. RMSE is calculated in the dimensionless response \(Y\). The constrained low-\(\Rey_{L_s}\) fit fixes the exponent at the theoretical value \(-1/3\); the constant plateau fit fixes \(p=0\).}
\label{tab:branchfits}
\centering
\begin{ruledtabular}
\begin{tabular}{lccccc}
Fit & \(N\) & \(C\) & \(p\) & \(R^2\) & RMSE \\
\hline
\(\Rey_{L_s}<10\), free power law
& 142
& \(2.038\)
& \(-0.322\)
& 0.909 & 0.449 \\
\(\Rey_{L_s}<10\), \(p=-1/3\)
& 142
& \(1.990\)
& \(-1/3\) (fixed)
& 0.908 & 0.450 \\
\(\Rey_{L_s}\ge10\), free power law
& 60
& \(0.98\)
& \(-0.048\)
& 0.157 & 0.085 \\
\(\Rey_{L_s}\ge10\), constant plateau
& 60
& \(0.866\)
& \(0\) (fixed)
& -- & 0.092 \\
\end{tabular}
\end{ruledtabular}
\end{table*}

The low-$\Rey_{L_s}$ result provides a direct quantitative test of the viscous prediction. The freely fitted exponent, $p=-0.323$, is in close agreement with the theoretical value of $-1/3$. Moreover, fixing $p=-1/3$ changes $R^2$ only from 0.909 to 0.908 and RMSE only from 0.449 to 0.450. Thus, the theoretically predicted $-1/3$ scaling is strongly supported by the data without being imposed by the fit.

For $\Rey_{L_s}\ge10$, the fitted exponent decreases in magnitude to $p=-0.025$, approximately one order of magnitude smaller than the low-$\Rey_{L_s}$ exponent. The small $R^2$ in this region should not be interpreted in the same way as for the sloped branch because the response itself varies only weakly with $\Rey_{L_s}$; RMSE is therefore more informative for assessing the near-plateau behavior. A constant model gives $Y=0.866$ with an RMSE of 0.092, only modestly larger than the RMSE of 0.085 obtained when a residual slope is allowed. The high-$\Rey_{L_s}$ branch is therefore only weakly dependent on Reynolds number rather than perfectly horizontal, consistent with the approach to an inertial asymptote.

The branch inference is also insensitive to the precise location of the crossover. Moving the cutoff between \(\Rey_{L_s}=5\) and 20 changes the low-\(\Rey_{L_s}\) exponent only from \(-0.3216\) to \(-0.3218\) and the high-\(\Rey_{L_s}\) exponent only from \(-0.0480\) to \(-0.0478\). The two asymptotic slopes are therefore not a consequence of tuning the cutoff at \(\Rey_{L_s}=10\).

Most importantly, in view of the shared occurrence of $Q$ in the normalized axes, the same branches were tested directly in dimensional variables. Through-origin regressions using the scales defined in Eqs.~(\ref{eq:SI}) and (\ref{eq:SV}) give
\begin{eqnarray}
S_c &=& 2.028\,S_V,
R^2=0.916,
{\rm RMSE}=1.11~{\rm mm},
\label{eq:dimfit_visc}\\
S_c &=& 0.840\,S_I,
R^2=0.930,
{\rm RMSE}=1.34~{\rm mm},
\label{eq:dimfit_inert}
\end{eqnarray}
for the $\Rey_{L_s}<10$ and $\Rey_{L_s}\ge10$ subsets, respectively. These dimensional fits retain the same physical separation between the viscous and inertial scales while using the measured $S_c$ itself as the response. They therefore provide an independent check on the interpretation of Fig.~\ref{fig:moodycollapse} beyond the normalized graphical collapse.

The quantitative results in Table~\ref{tab:branchfits} sharpen the regime-spanning interpretation of Fig.~\ref{fig:moodycollapse}. At large \(\Rey_{L_s}\), the normalized response becomes only weakly Reynolds-number dependent and is described nearly as well by a constant as by a free power law. As \(\Rey_{L_s}\) decreases, the response departs systematically from this near-plateau and follows an exponent statistically consistent with \(-1/3\). The close agreement between the free and theory-constrained low-\(\Rey_{L_s}\) fits shows that the observed slope is not being produced by forcing the theoretical exponent. Together with the dimensional fits in Eqs.~(\ref{eq:dimfit_visc}) and (\ref{eq:dimfit_inert}), this supports interpretation of the two branches as asymptotic manifestations of viscous and inertial balances rather than as unrelated empirical correlations.

It is also important to interpret how surface tension enters this framework. The master-curve organization does not imply that surface tension is negligible. Over the parameter range represented here, the data are consistent with capillarity entering the leading geometric organization through the capillary length \(l_c\) embedded in \(L_s=\max(l_c,a)\), while viscosity controls the magnitude of the low-\(\Rey_{L_s}\) correction. The present experiments contain two nominal interfacial-tension levels, and the literature points broaden the capillary-length range, but these datasets do not constitute an independent controlled sweep of \(\gamma\). The collapse therefore supports incorporation of capillarity through \(l_c\); it does not exclude smaller residual capillary effects, which remain a natural source of scatter about the master trend.

\subsection{Physical interpretation}
Equation~(\ref{eq:master}) admits a direct mechanistic interpretation.  
The inertial normalization $S_I$ reflects the requirement that suction-driven inertia overcome hydrostatic restoration to produce the deformation needed for entrainment.  
This baseline remains meaningful across the dataset because inertia and gravity remain present even when viscous stresses are significant.  
Viscous and shear effects therefore do not replace the inertial scale; instead, they shift the threshold by requiring a larger normalized height $Y$ at a given $Q$ to achieve comparable interfacial localization and curvature.  
In this sense, the rise of $Y$ at decreasing $\Rey_{L_s}$ quantifies how viscous resistance delays entrainment relative to the inertial baseline.

The exponent $-1/3$ follows from the viscous correction scale $S_V$ in Eq.~(\ref{eq:SV}), which arises from balancing viscous stresses against restoration effects evaluated over the scaling length $L_s$.  
Accordingly, the master-curve shape is interpretable in terms of the relative magnitudes of inertial and viscous stresses.  
The plateau corresponds to conditions where viscous stresses are small compared with inertia on the relevant length scale.  
The rising branch corresponds to conditions where viscous resistance becomes comparable and then increasingly influential as $\Rey_{L_s}$ decreases.

A recurring difficulty in regime-spanning descriptions is selecting an appropriate horizontal scaling length. Using \(L_s=a\) alone omits gravity--capillary curvature when \(l_c>a\), whereas using \(L_s=l_c\) alone omits the finite near-nozzle scale when \(a>l_c\). The composite definition \(L_s=\max(l_c,a)\) is the minimal prescription that recovers both geometric limits without introducing an additional fitted parameter. The fact that both orderings occur within the compiled literature data gives this choice an empirical test across the collapse. It should nevertheless be regarded as a practical crossover scale rather than a uniquely derived smooth interpolation; alternative smooth crossover functions may be examined when denser data are available near \(a\approx l_c\).

\subsection{Non-Newtonian collapse}
The placement of the shear-thinning xanthan-gum cases on the same primary axes supports a specific, but deliberately limited, interpretation of non-Newtonian effects in selective withdrawal. Over the conditions examined, the data do not require a separate entrainment mechanism. Instead, lower-layer rheology modifies the viscous stress contribution through the shear-rate-dependent factor \(\Lambda_{\mathrm{NN}}\) in Eq.~(\ref{eq:LambdaNN}), or equivalently through the local effective viscosity in Eq.~(\ref{eq:etaeff}).

The estimate of \(\dot{\gamma}_{\mathrm{eff}}\) is necessarily an order-of-magnitude representation of a spatially varying flow near the hump apex. Its sensitivity can be quantified directly from the power-law rheology. If the estimated shear rate is multiplied by a factor \(c\), then
\begin{equation}
\frac{\eta_{L,\mathrm{eff}}(c\dot{\gamma})}
{\eta_{L,\mathrm{eff}}(\dot{\gamma})}
=c^{\,n-1}.
\label{eq:rheosensitivity}
\end{equation}
For the measured \(n=0.62\)--0.64, changing the characteristic shear rate by a factor of 2 changes the apparent viscosity by a factor of approximately 0.77--0.78, while a factor-of-10 change gives approximately 0.42--0.44. Thus, uncertainty in the local shear-rate estimate can materially alter the magnitude of the lower-layer viscous correction, but it does not change the inertial normalization or the observed \(-1/3\) branch of the primary collapse. The local flow also contains extension and strong spatial gradients that are not resolved by the scalar power-law estimate. The present treatment should therefore be read as a leading generalized-Newtonian correction, with local velocimetry, CFD, or resolved hump geometry needed for a more complete rheological description.

\subsection{Scope and limitations}
The new experiments use a single withdrawal-tube diameter, so geometric generality does not come from the present apparatus alone. The literature datasets broaden the tube-size and density-difference ranges and, importantly, populate both \(l_c>a\) and \(a>l_c\) orderings, but the compiled data remain sparser in the crossover region than in the asymptotic branches. The threshold \(\Rey_{L_s}\approx10\) should therefore be interpreted as an empirical crossover scale for the present compilation, not as a universal sharp transition. Likewise, the present evidence supports the master-curve organization for the Newtonian and shear-thinning systems examined here; broader generalized-Newtonian behavior, controlled variation of interfacial tension, and denser variation of tube size remain tests of the framework rather than established consequences of it.

\section{Conclusion}
Selective-withdrawal experiments spanning viscocapillary, transitional, and inertia-influenced operating conditions were analyzed together with representative literature data. A regime-spanning organization emerges when the critical submergence height is normalized by the inertial scale, $S_I\sim(Q^2/g)^{1/5}$, and expressed in terms of the Reynolds-like control parameter $\Rey_{L_s}$ constructed using $L_s=\max(l_c,a)$. The resulting Moody-diagram representation is consistent with the composite form
\begin{eqnarray}
Y = \frac{S_c g^{1/5}}{Q^{2/5}}
= A + B\,\Rey_{L_s}^{-1/3},
\label{eq:concl_master}
\end{eqnarray}
where the additive form is interpreted as a compact interpolation between the inertial and viscous asymptotes rather than as an exact solution of the full free-boundary problem. In this representation, the high-$\Rey_{L_s}$ regime approaches an inertia-dominated plateau, whereas decreasing $\Rey_{L_s}$ produces a systematic increase in the normalized critical height associated with growing viscous resistance.

The central scaling behavior is supported quantitatively rather than by visual collapse alone. For $\Rey_{L_s}<10$, the free regression gives $p=-0.322$ with $R^2=0.909$, in close agreement with the theoretically predicted exponent of $-1/3$. Fixing the exponent at $p=-1/3$ changes the fit quality only negligibly, showing that the predicted viscous scaling is supported by the data rather than imposed through the regression. For $\Rey_{L_s}\ge10$, the exponent decreases strongly in magnitude to $-0.048$, while a constant plateau provides a comparably accurate description of the data. This marked reduction in slope is consistent with the transition toward an inertia-dominated regime in which the normalized critical height becomes only weakly dependent on $\Rey_{L_s}$. The inferred branch behavior is also insensitive to reasonable changes in the crossover location around $\Rey_{L_s}=10$.

Because $Q$ appears in both normalized axes, the asymptotic behavior was also evaluated directly using dimensional variables. In the viscous branch, the measured critical height follows
\begin{equation}
S_c=2.028\,S_V,
\qquad R^2=0.916,
\end{equation}
where
\begin{equation}
S_V\sim
\left(
\frac{\mu_w Q}{\Delta\rho gL_s}
\right)^{1/3}.
\end{equation}
In the inertia-dominated branch,
\begin{equation}
S_c=0.840\,S_I,
\qquad R^2=0.930.
\end{equation}
The agreement of these dimensional regressions with the two branches of the normalized representation provides an important independent validation of the proposed scaling picture. In particular, it shows that the viscous and inertial balances remain predictive when the measured $S_c$ itself is used as the response, rather than relying solely on a nondimensional plot in which $Q$ appears on both axes.

Surface tension remains an essential part of the framework through the capillary length,
\begin{equation}
l_c=\sqrt{\frac{\gamma}{\Delta\rho g}},
\end{equation}
which contributes to the characteristic lateral scale $L_s$. The use of $L_s=\max(l_c,a)$ provides a practical way to connect conditions in which either gravity--capillary curvature or the withdrawal-tube geometry sets the relevant lateral scale. Thus, capillarity is retained in the geometric organization of the problem even though the principal transition in the master curve reflects the changing relative importance of viscous and inertial effects.

For shear-thinning lower layers, the results indicate that non-Newtonian effects enter through the shear-rate-dependent viscous contribution, represented by $\eta_{L,\text{eff}}$ and $\Lambda_{\mathrm{NN}}$. In this framework, shear thinning modifies the viscous correction rather than altering the inertial baseline that anchors the normalization. This explains why Newtonian and shear-thinning cases can be compared on the same primary axes while still allowing measurable differences in the entrainment threshold. Improved estimates of the local interfacial shear-rate scale can further refine the evaluation of $\eta_{L,\text{eff}}$, while any remaining surface-tension or geometric dependence can be examined as secondary structure about the primary collapse.

Taken together, the results support a physical picture in which inertia and gravity establish the baseline entrainment scale, while viscous resistance provides an increasingly important correction as $\Rey_{L_s}$ decreases. The transition between these limits can therefore be represented within a single physically interpretable framework rather than by switching between separate viscocapillary and inertial correlations. The agreement between the measured and predicted low-$\Rey_{L_s}$ exponent, the emergence of a weakly varying high-$\Rey_{L_s}$ plateau, and the independent dimensional validation collectively strengthen the interpretation of the proposed master curve as a quantitative bridge between historically separated selective-withdrawal regimes. More broadly, the Moody-diagram-type representation provides a compact framework for comparing entrainment thresholds across the Newtonian and shear-thinning systems examined here while retaining clear connections to the underlying balances of inertia, viscosity, gravity, and capillarity.

\bibliographystyle{unsrt}
\bibliography{apssamp}

\end{document}